\documentclass[aps, prd, showkeys,superscriptaddress, nofootinbib, floatfix]{revtex4-2}

\usepackage{amsmath}
\usepackage{graphicx}
\usepackage{dcolumn}
\usepackage{bm}
\usepackage{amsfonts}
\usepackage{mathrsfs}
\usepackage{amssymb}
\usepackage{slashed}
\usepackage[colorlinks=true,linkcolor=blue]{hyperref}
\usepackage[capitalize]{cleveref}
\usepackage[x11names]{xcolor}
\usepackage{bm}
\usepackage{empheq}
\usepackage{ragged2e}
\usepackage[normalem]{ulem}

\usepackage{enumitem}

\crefname{section}{Sec.}{Sec.}
\Crefname{section}{Section}{Sections}

\usepackage[super]{nth}

\usepackage{dsfont}
\usepackage{url}
\usepackage{lipsum}
\usepackage{color}
\usepackage{colortbl}

\def\Eq#1{Eq.~\eqref{#1}}

\begin{document}

\title{Covariant diffusion tensor for jet momentum broadening out of equilibrium}

\author{Isabella Danhoni}
\email{idanhoni@illinois.edu}
\affiliation{Illinois Center for Advanced Studies of the Universe, Department of Physics, University of Illinois at
Urbana-Champaign, Urbana, IL 61801, USA}
\author{Nicki Mullins}
\email{nmmulli2@ncsu.edu}
\affiliation{Department of Physics, North Carolina State University, Raleigh, NC 27695, USA}
\author{Jorge Noronha}
\email{jn0508@illinois.edu}
\affiliation{Illinois Center for Advanced Studies of the Universe, Department of Physics, University of Illinois at
Urbana-Champaign, Urbana, IL 61801, USA}

\begin{abstract}
Jets are produced in the earliest stages of heavy-ion collisions, where they can interact with a medium that is not yet close to local equilibrium. 
Motivated by this, we generalize the usual jet transport coefficient $\hat q$ to a Lorentz-covariant diffusion tensor $\hat q^{\mu\nu}$ within a leading-order elastic (Boltzmann/Fokker--Planck) description of jet--medium interactions. 
The tensor formulation organizes medium effects in a frame-covariant way and reveals additional information beyond the standard scalar definition, including energy diffusion and off-diagonal components that encode correlations between energy and momentum exchange which are absent (or redundant) in equilibrium.
We illustrate the formalism in (tree-level) massless $\lambda\varphi^4$ theory for isotropic but out-of-equilibrium states. 
For sufficiently large jet momentum, quantum statistical effects become subleading, so that the non-equilibrium evolution can be studied reliably in the classical (Boltzmann) limit. 
This allows us to solve the corresponding Boltzmann equation for the medium and determine the time dependence of $\hat q^{\mu\nu}$ as the system approaches equilibrium. 
We find that out-of-equilibrium corrections can either enhance or reduce jet momentum broadening, depending on the initial distribution function.
\end{abstract}

\maketitle

\tableofcontents

\section{Introduction}
\label{Sec:Introduction}

In the earliest stages of ultrarelativistic heavy-ion collisions, hard scatterings produce highly energetic partons that subsequently traverse the evolving medium, undergoing medium-induced energy loss and momentum broadening before fragmenting into jets. Measurements of these modifications provide some of the sharpest probes of the quark--gluon plasma \cite{Connors:2017ptx}, and they motivate a sustained effort to develop quantitative theoretical descriptions of jet quenching and confront them with experimental data \cite{Majumder:2010qh,Cao:2024pxc,Mehtar-Tani:2025rty}.

In the case where only elastic processes are considered, the medium modification of a hard parton is commonly characterized by the jet quenching parameter $\hat q$, defined as the transverse-momentum diffusion rate
\begin{equation}
\hat q \equiv \frac{d}{dL}\,\langle p_\perp^2\rangle,
\label{eq:qhat_def_intro}
\end{equation}
i.e.\ the rate of transverse momentum broadening accumulated per unit path length $L$, where $p_\perp$ denotes momentum transverse to the jet direction in the usual local-rest-frame setup
\cite{Baier:1994bd,Baier:1996sk,Baier:1996kr} (see also \cite{Gyulassy:1993hr} for early multiple-scattering treatments that underpin the description of this quantity).

Since hard partons are produced in the initial hard scatterings, they begin propagating through the medium essentially immediately after the nuclear impact, i.e.\ well before the system has hydrodynamized into a nearly equilibrated quark--gluon plasma. This has motivated a number of efforts to quantify jet broadening and energy loss during the pre-equilibrium evolution using various frameworks, see e.g.\ \cite{Citron:2018lsq, Ipp:2020mjc, Ipp:2020nfu, Carrington:2021dvw, Carrington:2022bnv, Avramescu:2023qvv, Andres:2022bql, Boguslavski:2023alu,Hauksson:2021okc, Hauksson:2023tze,Barata:2025zku,Barata:2024xwy,Kuzmin:2023hko,Barata:2023zqg,Barata:2023qds,Barata:2022krd,Barata:2021wuf,Barata:2022utc,Altenburger:2025iqa,He:2015pra,Renk:2006sx,Mehtar-Tani:2022zwf,Salgado:2003gb}.  
During these earliest times the medium can be far from equilibrium, with large momentum-space anisotropies and rapidly evolving collective flow. Correspondingly, its energy--momentum tensor need not be close to the ideal form and may receive non-negligible contributions from dissipative stresses (e.g.\ the shear-stress tensor $\pi^{\mu\nu}$ and the bulk pressure $\Pi$). 
In such conditions there is, in general, no symmetry reason for momentum broadening to be characterized by a single scalar: the broadening process is intrinsically direction dependent and can involve nontrivial energy--momentum correlations that are absent (or redundant) in equilibrium.
Equivalently, the scalar $\hat q$ commonly used in equilibrated media captures only a restricted projection of a more general diffusion structure; applying an equilibrium-based scalar characterization \cite{Baier:1996kr} to far-from-equilibrium stages can therefore miss the tensorial information associated with anisotropy and flow. For that reason, there has been as series of new developments in jet quenching calculations, both in the transverse direction to the jet axis
and along the longitudinal direction, see for example~\cite{Barata:2025wnp,Barata:2025htx,Antiporda:2021hpk}. 
In particular, out of equilibrium the notion of a unique ``transverse plane'' tied to a local rest frame is, at best, approximate, and different projections of the momentum transfer need not be equivalent.
This highlights the need for introducing a Lorentz-covariant tensorial generalization of $\hat q$ that remains well-defined in flowing and far-from-equilibrium media and makes explicit the additional diffusion information allowed by reduced symmetry.
In this context, kinetic theory provides a natural framework to treat such out-of-equilibrium media while remaining consistent with macroscopic conservation laws~\cite{Heinz:1984yq, Bass:1998ca, Arnold:2000dr, Molnar:2001ux, Xu:2004mz, Rocha:2023ilf}. In fact, kinetic theory has recently been used to assess jet--medium interactions out of equilibrium in Refs.\ \cite{Schlichting:2020lef,Boguslavski:2023waw,Boguslavski:2023alu,Barata:2025agq}.

In this work, we generalize the usual jet transport coefficient $\hat q$ to a Lorentz-covariant diffusion tensor $\hat q^{\mu\nu}$ within a leading-order elastic (Boltzmann/Fokker--Planck) description of jet--medium interactions in a generic, possibly time-dependent medium.
The tensorial formulation provides a frame-covariant organization of medium effects through a decomposition into scalar functions built from the available four-vectors and tensors, and it naturally introduces off-diagonal components that encode correlations between energy and momentum exchange, as well as an energy-diffusion sector that is invisible in the standard scalar definition in equilibrium.
We discuss the physical interpretation of each component by relating $\hat q^{\mu\nu}$ to the growth rates of appropriate cumulants of the exchanged four-momentum, and  derive general constraints that follow from its interpretation as a diffusion tensor in a Fokker--Planck description.

In this work, we investigate $\hat q^{\mu\nu}$ explicitly in a simplified setting that allows for analytic control, namely tree-level massless $\lambda\varphi^4$ theory, considering isotropic but out-of-equilibrium medium states that relax toward equilibrium. 
By comparing Bose--Einstein (quantum) and Boltzmann (classical) medium statistics, we demonstrate that for sufficiently large jet momentum quantum statistical effects become subleading, so that the dynamics relevant for $\hat q^{\mu\nu}$ can be reliably described in the classical Boltzmann limit. 
We then evaluate $\hat q^{\mu\nu}(t)$ out of equilibrium by solving the corresponding Boltzmann equation for the medium in this classical setting. We show that the sign and magnitude of the non-equilibrium corrections depend sensitively on the initial distribution function, leading to either enhanced or reduced momentum broadening relative to equilibrium.

This paper is organized as follows. In Sec.~\ref{Sec:kinetic_theory} we briefly review the kinetic-theory setup and introduce the Lorentz-covariant diffusion tensor $\hat q^{\mu\nu}$, which encodes the momentum-diffusion information obtainable from a leading-order Boltzmann equation with elastic interactions. In Sec.~\ref{Sec: Fokker_Planck} we motivate the tensorial structure using a Fokker--Planck description and discuss the physical interpretation of the different components of $\hat q^{\mu\nu}$. In Sec.~\ref{Sec:phi^4} we compute $\hat q^{\mu\nu}$ in tree-level massless $\lambda\varphi^4$ theory, compare results obtained with Bose--Einstein (quantum) and Boltzmann (classical) medium statistics, and study their large-momentum behavior. In Sec.~\ref{sec:outofeq} we evaluate the time dependence $\hat q^{\mu\nu}(t)$ for explicit isotropic out-of-equilibrium distributions describing relaxation toward equilibrium, obtained by solving the classical Boltzmann equation within a method-of-moments framework, and compare the resulting broadening to the equilibrium limit. Finally, in Sec.~\ref{Sec:Conclusions} we summarize our results and provide an outlook.

Notation: In the present work, we employ the mostly minus $(+,-,-,-)$ metric signature, and natural units so that $\hbar = c = k_{B} = 1$. The fluid 4-velocity is $u^\mu$, and we also use the orthogonal projector, $\Delta^{\mu\nu} \equiv g^{\mu\nu} - u^\mu u^\nu$ (which projects any tensor onto the space orthogonal to $u^\mu$). When convenient, we denote the scalar product between four-vectors as follows $u_\mu K^\mu = u \cdot K$. We use $\mathbf{k}$ as a notation for 3-dimensional vectors, with the exception of $\vec{q}_\perp$ (representing transverse exchanged momentum vectors).
\section{Kinetic theory and a covariant definition of jet momentum broadening}
\label{Sec:kinetic_theory}
\begin{figure}
    \centering
    \includegraphics[width=0.35\linewidth]{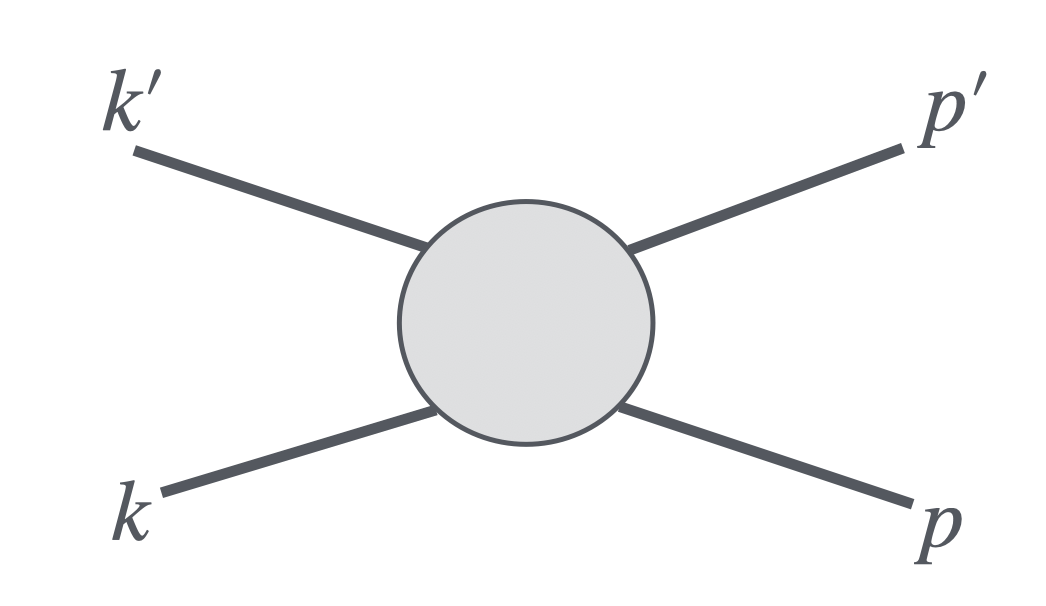}
    \caption{Schematic view of the elastic $2\leftrightarrow 2$ scattering process described by the Boltzmann equation.}
    \label{fig:collision_diagram}
\end{figure}

In this section we introduce the kinetic-theory framework used throughout this work. We start from the standard scalar jet-quenching parameter $\hat q$, which will serve as a baseline for the Lorentz-covariant tensorial definition employed in the rest of the paper. We describe the medium in terms of single-particle distribution functions $f(x^{\mu},\mathbf{p})\equiv f_{\mathbf{p}}$, whose evolution is governed by the relativistic Boltzmann equation (see, e.g., \cite{DeGroot:1980dk}). The jet is represented by an on-shell four-momentum $K^\mu$ propagating through this medium. For a general collision operator, the Boltzmann equation reads
\begin{equation}
\label{eq:Boltzmann}
K^{\mu} \partial_{\mu} f_{\mathbf{k}}  
= \mathcal{C}[f_{\mathbf{k}}] \, . 
\end{equation}

Here we label ingoing particle momenta as $K^\mu$ (the jet) and $K'^\mu$, and outgoing momenta as $P^\mu$ and $P'^\mu$, as shown in Fig.~\ref{fig:collision_diagram}. In this work we define the 3-momentum transferred in each elastic scattering as $\mathbf{q}\equiv \mathbf{k}-\mathbf{p}$. Since the explicit form of the differential cross section is only required for the specific computations presented later, we defer its specification to the relevant section. The definitions introduced here are otherwise general and independent of the microscopic interaction.

We take $\mathcal{C}$ to be the collision operator acting on the jet distribution. For the purpose of jet momentum broadening, we restrict to elastic $2\leftrightarrow 2$ scattering processes, so that the collision operator is
\begin{equation}
\label{eq:2to2}
   \mathcal{C}_a[f_{\mathbf{k}}] =
\frac{1}{2}\int dP \, dK' \, dP' \, \mathcal{W}_{a} \Big( f_{\mathbf{k}}f_{\mathbf{k'}} \tilde{f}_{\mathbf{p}}\tilde{f}_{\mathbf{p'}}-  f_{\mathbf{p'}}f_{\mathbf{p}}\tilde{f}_{\mathbf{k'}} \tilde{f}_{\mathbf{k}}\Big),
\end{equation}
where $\mathcal{W}_a=\mathcal{W}_{pp' \leftrightarrow kk'}$ denotes the relevant scattering kernel (rate) for the process. The Lorentz-invariant integral measure for on-shell massless particles is $dK \equiv d^{3}\mathbf{k}/[(2\pi)^3 2k^{0}]$, and we use $\tilde{f}_{\mathbf{k}} \equiv (1+a f_{\mathbf{k}})$, with $a=1$ for bosons, $a=-1$ for fermions, and $a=0$ for classical (Boltzmann) statistics. Furthermore, note that $\mathcal{W}_a\ge 0$ and the on-shell constraints are incorporated in its definition.

In this framework, the jet-quenching parameter $\hat{q}$ quantifies transverse momentum broadening due to elastic scattering, and can be defined in terms of the elastic scattering rate \cite{Arnold:2008vd, Caron-Huot:2008zna,Ghiglieri:2015ala} as
\begin{equation}
    \hat{q} = \int d^2\vec{q}_\perp \, \vec{q}_\perp^{\,2} \frac{d\Gamma_{a,{\rm el}}}{d^2\vec{q}_\perp} \, ,
    \label{q_hat}
\end{equation}
where $\vec{q}_\perp$ is the momentum exchanged transverse to the jet direction (in the standard local-rest-frame setup), and $\Gamma_{a,{\rm el}}$ denotes the elastic scattering rate.

The collision operator can be written in terms of gain and loss contributions,
\begin{equation}
\begin{aligned}
    \mathcal{C}_{a,{\rm gain}}[f_{\mathbf{k}}] &=  \frac{1}{2}\int dP \, dK' \, dP' \, \mathcal{W}_{a} \, (f_{\mathbf{p'}}f_{\mathbf{p}}\tilde{f}_{\mathbf{k'}} \tilde{f}_{\mathbf{k}})\, ,\\
    \mathcal{C}_{a,{\rm loss}}[f_{\mathbf{k}}] &=  \frac{1}{2}\int dP \, dK' \, dP' \, \mathcal{W}_{a} \, (f_{\mathbf{k}}f_{\mathbf{k'}} \tilde{f}_{\mathbf{p}}\tilde{f}_{\mathbf{p'}})\, .
\end{aligned}
\label{eq:loss_term}
\end{equation}
With these definitions, the elastic scattering rate can be related directly to the loss term as
\begin{equation}
\label{Eq:elastic}
    \Gamma_{a,{\rm el}}[\mathbf{k}]=\frac{1}{2}\int dP \, dK' \, dP'\,\mathcal{W}_a \, f_{\mathbf{k'}} \tilde{f}_{\mathbf{p}}\tilde{f}_{\mathbf{p'}}\, .    
\end{equation}
Inserting \Eq{Eq:elastic} into \Eq{q_hat}, one finds
\begin{equation}
\label{Eq:qhat_scalar}
    \hat{q}[\mathbf{k}]=\frac{1}{2|\mathbf{k}|}\int dP \, dK' \, dP' \,\vec{q}_\perp^{\,2}\,\mathcal{W}_{a}\,f_{\mathbf{k'}} \tilde{f}_{\mathbf{p}}\tilde{f}_{\mathbf{p'}}\, .
\end{equation}

We emphasize that the $\hat{q}$ in \Eq{Eq:qhat_scalar} is intrinsically tied to a specific frame, the local rest frame (LRF) of the medium, and to a specific choice of transverse plane. However, in realistic heavy-ion applications the medium is neither static nor isotropic: it has spacetime-dependent collective flow and is generally out of equilibrium, especially at early times. In such situations it is natural to introduce the exchanged four-momentum
\begin{equation}
q^\mu \equiv (K-P)^\mu,
\end{equation}
and to generalize \Eq{Eq:qhat_scalar} to a Lorentz-covariant tensor. Therefore, we define
\begin{equation}
\label{Eq:qhat_tensor}
    \hat{q}^{\mu\nu}(K)=\frac{1}{2\,u_\alpha K^\alpha}\int dP \, dK' \, dP'\, q^\mu q^\nu\,\mathcal{W}_{a}\,f_{\mathbf{k'}} \tilde{f}_{\mathbf{p}}\tilde{f}_{\mathbf{p'}}\, ,
\end{equation}
where $u^\mu$ is the local fluid four-velocity, so that $u_\alpha K^\alpha$ is the jet energy in the local rest frame. Within this formulation, medium-induced broadening is described by a symmetric rank-2 tensor $\hat{q}^{\mu\nu}$, which encodes how the variance of the jet four-momentum accumulates along different directions. The tensor depends on the jet four-momentum $K^\mu$ and on the local medium data (at minimum $u^\mu$, and, out of equilibrium, potentially additional tensors such as $\pi^{\mu\nu}$), while transforming covariantly under Lorentz transformations. This quantity admits a natural interpretation as a momentum-space diffusion matrix as contractions with vectors specify diffusion along the corresponding directions.

From a theoretical perspective, the tensorial definition \Eq{Eq:qhat_tensor} has several advantages. First, it provides a covariant description of jet momentum broadening that can be evaluated in any frame, which is essential in the presence of spacetime-dependent flow and along arbitrary jet trajectories. Second, it naturally accommodates anisotropies and non-equilibrium effects in the medium, allowing one to distinguish, for example, between broadening along and across principal directions selected by the local stress tensor. In what follows, we connect this tensorial formulation to a Fokker--Planck picture that clarifies the interpretation of $\hat{q}^{\mu\nu}$ and its individual components.

\section{Covariant jet momentum broadening as a diffusion tensor}
\label{Sec: Fokker_Planck}

In this section, we motivate the tensorial structure of $\hat q$ starting from a Fokker--Planck formulation. This approach provides a direct physical interpretation of $\hat{q}^{\mu\nu}$ as a momentum-space diffusion tensor; for other discussions on this topic we refer the reader to~\cite{Baier:1996kr,Baier:1996sk,Zakharov:1997uu,Arnold:2000dr,Arnold:2003zc,Majumder:2007hx,Carrington:2021dvw}. Covariant diffusion tensors and relativistic Fokker--Planck descriptions have previously appeared in~\cite{Moore:2004tg,Denicol:2012cn,Ghiglieri:2018dib,Rajagopal:2025rxr}, though not in the context of jet momentum broadening considered here. 

We begin with the assumption that the momentum transfer $q^\mu$ is small compared to the jet momentum $K^\mu$, so that the right-hand side of Eq.~\eqref{eq:2to2} can be expanded in powers of $q^\mu$, leading to a Fokker--Planck approximation of the collision term. Expanding the distribution function $f_{\mathbf{k} + \mathbf{q}}$ around $K^\alpha$ yields
\begin{equation}
f_{\mathbf{k} + \mathbf{q}}=f_{\mathbf{k}}
+ q^\alpha \frac{\partial f_{\mathbf{k}}}{\partial K^\alpha}
+ \frac{1}{2} q^\alpha q^\beta
\frac{\partial^2 f_{\mathbf{k}}}{\partial K^\alpha \partial K^\beta}
+ \cdots \, .
\label{eq:FPexpansion}
\end{equation}
We note that this expression should be understood in terms of its components in a physically preferred local frame. In particular, given the natural timelike vector $u^\mu$, we introduce an orthonormal tetrad $e^\mu_{(a)}$ with $e^\mu_{(0)}=u^\mu$, and decompose
\begin{equation}
    q^\mu = q_{(a)}e^\mu_{(a)}\, ,\hspace{1cm} K^\mu = K_{(a)}e^\mu_{(a)}\, .
\end{equation}
Throughout this section, derivatives with respect to $K^\alpha$ are understood as derivatives along the independent momentum components in this local frame (i.e.\ tangential to the mass shell), so that the manipulations below are performed on the on-shell distribution.

Here, the small--angle (or soft--scattering) assumption means that the kernel $W_a(K,q)$ has support only for momentum transfers whose components satisfy
\begin{equation}
    |q_{(a)}|\ll|K_{(0)}|\, .
\end{equation}
In particular, for an ultrarelativistic jet, $K_{(0)}$ is of the order of the jet energy in the LRF. In this sense, $q^\mu$ is ``small'' even though it need not be timelike and may be spacelike. Equivalently, $W_a(\mathbf{k},\mathbf{q})$ is sharply peaked around $q^\mu=0$ on a momentum scale much smaller than the scale over which $f_{\mathbf{k}}$ varies in $K^\alpha$. This separation of scales justifies the Taylor expansion of $f_{\mathbf{k+q}}$ in powers of $q^\mu$.

Substituting \Eq{eq:FPexpansion} into \Eq{eq:2to2} and retaining terms up to second order in $q$ one obtains
\begin{equation}
\begin{aligned}
\mathcal{C}_a[f_{\mathbf{k}}]
&\simeq \int d^4 q \, W_a(K,q)
\left[
q^\alpha \frac{\partial f_{\mathbf{k}}}{\partial K^\alpha}
+ \frac{1}{2} q^\alpha q^\beta
\frac{\partial^2 f_{\mathbf{k}}}{\partial K^\alpha \partial K^\beta}
\right] \\
&= \frac{\partial}{\partial K^\alpha}
\left[
A_a^\alpha(K)\, f_{\mathbf{k}}
\right]
+ \frac{1}{2}
\frac{\partial^2}{\partial K^\alpha \partial K^\beta}
\left[
B_a^{\alpha\beta}(K)\, f_{\mathbf{k}}
\right] .
\end{aligned}
\end{equation}
where, up to surface terms in momentum space (which vanish under suitable boundary conditions), we have defined the \emph{drift vector} $A_a^{\alpha}(K)$ and the symmetric \emph{diffusion tensor} $B_a^{\alpha\beta}(K)$ by
\begin{equation}
A_a^{\alpha}(K) \equiv \int d^{4}q \, q^{\alpha} \, W_a(K,q),
\qquad
B_a^{\alpha\beta}(K) \equiv \int d^{4}q \, q^{\alpha} q^{\beta} \, W_a(K,q).
\end{equation}
We emphasize to the reader that the symmetric tensor $B_a^{\alpha\beta}(K)$ therefore plays the role of a momentum-space diffusion matrix for the jet, which will be analyzed in more detail in what follows.

Inserting the Fokker--Planck approximation for the collision term derived above into Eq.~\eqref{eq:Boltzmann} yields the relativistic generalization of the ordinary Fokker--Planck equation describing a particle undergoing stochastic kicks in momentum space:
\begin{equation}
\label{eq:rel_fokkerpl}
    K^{\mu} \partial_{\mu} f_{\mathbf{k}} = \frac{\partial}{\partial K^\alpha}
\left[
A_a^\alpha(K)\, f_{\mathbf{k}}
\right]
+ \frac{1}{2}
\frac{\partial^2}{\partial K^\alpha \partial K^\beta}
\left[
B_a^{\alpha\beta}(K)\, f_{\mathbf{k}}
\right] .
\end{equation}

To understand the physical meaning of $B^{\alpha\beta}$ (and hence of $\hat q^{\mu\nu}$), we parametrize the jet trajectory $x^\mu(\lambda)$ by an affine parameter $\lambda$ such that
\begin{equation}
\frac{dx^\mu}{d\lambda} = K^\mu,
\qquad
\frac{d}{d\lambda} f(\lambda; K)
= \frac{dx^\mu}{d\lambda} \, \partial_\mu f_{\mathbf{k}}
= K^\mu \partial_\mu f_{\mathbf{k}}\, .
\end{equation}
With this choice, \Eq{eq:rel_fokkerpl} becomes an evolution equation in $\lambda$ at fixed $K$:
\begin{equation}
\frac{d}{d\lambda} f_{\mathbf{k}}
=
\frac{\partial}{\partial K_\alpha}
\left[
A^\alpha(K)\, f_{\mathbf{k}}
\right]
+
\frac{1}{2}
\frac{\partial^2}{\partial K_\alpha \partial K_\beta}
\left[
B^{\alpha\beta}(K)\, f_{\mathbf{k}}
\right].
\end{equation}
Here $\lambda$ is a convenient evolution parameter along the jet worldline. In the fluid rest frame, $u^\mu = (1,0,0,0)$, and for an ultrarelativistic jet, $\lambda$ is proportional to the physical time $t$ and hence to the path length $\ell$ (up to an overall constant factor). The precise proportionality factor will not affect the structure of the moment equations below.

Next, we consider the evolution of the first and second moments of the jet momentum distribution with respect to $\lambda$:
\begin{equation}
\label{eq:param}
\langle K^\mu \rangle(\lambda)
\equiv \int dK \, K^\mu \, f_{\mathbf{k}}\, ,
\qquad
\langle K^\mu K^\nu \rangle(\lambda)
\equiv \int dK \, K^\mu K^\nu \, f_{\mathbf{k}}\, ,
\end{equation}
where the integral is over the on--shell measure. We focus on these moments because their evolution will allow us to identify $\hat{q}^{\mu\nu}$ with the growth rate of momentum fluctuations.

Differentiating $\langle K^\mu \rangle$ with respect to $\lambda$ and using \Eq{eq:param}, we find
\begin{equation}
\begin{aligned}
\frac{d}{d\lambda} \langle K^\mu \rangle
&= \int dK \, K^\mu \, \frac{d}{d\lambda} f_{\mathbf{k}} \\
&= \int dK \, K^\mu
\left\{
\frac{\partial}{\partial K_\alpha}
\left[
A^\alpha(K)\, f_{\mathbf{k}}
\right]
+ \frac{1}{2}
\frac{\partial^2}{\partial K_\alpha \partial K_\beta}
\left[ B^{\alpha\beta}(K)\, f_{\mathbf{k}}
\right] \right\}.
\end{aligned}
\end{equation}
Integrating by parts in momentum space and assuming that the products
$A^\alpha(K) f_{\mathbf{k}}$ and $B^{\alpha\beta}(K) f_{\mathbf{k}}$ vanish sufficiently fast at large momentum so that all boundary terms vanish, the contribution from the second-derivative term cancels identically, and we obtain
\begin{equation}
\frac{d}{d\lambda} \langle K^\mu \rangle
=
-\int dK \, A^\mu(K)\, f_{\mathbf{k}}
= -\langle A^\mu(K) \rangle \, .
\end{equation}
Similarly, for the second moment, performing the integrations by parts and using the standard multidimensional Fokker--Planck moment identities (the generalization of the 1D case), one finds
\begin{equation}
\frac{d}{d\lambda} \langle K^\mu K^\nu \rangle
= -
 \langle A^\mu(K)\, K^\nu + A^\nu(K)\, K^\mu \rangle
+ \langle B^{\mu\nu}(K) \rangle\,  .
\label{eq:deriv}
\end{equation}
This expression shows explicitly that the diffusion tensor $B^{\mu\nu}(K)$ governs the growth of momentum fluctuations.

It is then natural to decompose the second moment into a mean part and a covariance part. Define the covariance of the jet momentum by
\begin{equation}
\langle \Delta K^\mu \Delta K^\nu \rangle
\equiv
\langle K^\mu K^\nu \rangle
-
\langle K^\mu \rangle \langle K^\nu \rangle .
\end{equation}
Differentiating and using \Eq{eq:deriv} together with the evolution of $\langle K^\mu \rangle$, one finds that, in general, drift terms contribute to the evolution of the covariance through correlations between $A^\mu(K)$ and $K^\nu$. For a \emph{narrow jet distribution} in momentum space, such correlations are subleading, and we may approximate
\begin{equation}
\frac{d}{d\lambda}
\langle \Delta K^\mu \Delta K^\nu \rangle (\lambda)
\simeq
\langle B^{\mu\nu}(K) \rangle \, .
\end{equation}
Within this narrow--jet Fokker--Planck regime, the diffusion tensor $B^{\mu\nu}(K)$ controls the growth (per unit $\lambda$) of the momentum covariance. For a narrow jet distribution peaked around some momentum $K^\mu$, we may further approximate $B^{\mu\nu}(K)$ by its value at that momentum and pull it out of the average:
\begin{equation}
\frac{d}{d\lambda}
\langle \Delta K^\mu \Delta K^\nu \rangle (\lambda)
\simeq
B^{\mu\nu}(K)\, .
\end{equation}
These approximations describe highly energetic jets with narrow momentum distributions, as considered in our calculations.

The jet broadening tensor $\hat q^{\mu\nu}(K,u)$ is then defined by
\begin{equation}
E_k(K,u) \equiv  u_\alpha K^\alpha > 0,
\qquad
\hat q^{\mu\nu}(K,u)
\equiv
\frac{1}{E_k(K,u)}\, B^{\mu\nu}(K)
=
 \frac{1}{u \cdot K}\, B^{\mu\nu}(K).
\end{equation}
This normalization converts diffusion per unit affine parameter $\lambda$ into diffusion per unit physical path length in the local rest frame. Indeed, in the fluid rest frame,
\begin{equation}
\frac{dt}{d\lambda}=\frac{dx^0}{d\lambda}=K^0=E_k
\qquad\Rightarrow\qquad
\frac{d}{dt}=\frac{1}{E_k}\frac{d}{d\lambda},
\end{equation}
and for an ultrarelativistic jet one may identify $d\ell \simeq dt$. Therefore,
\begin{equation}
\frac{d}{d\ell}
\langle \Delta K^\mu \Delta K^\nu \rangle
\simeq
\hat q^{\mu\nu}(K,u),
\qquad
\text{LRF \,\, (narrow high--energy jet)}.
\label{eq:cov_evol}
\end{equation}
Thus, one can see that $\hat q^{\mu\nu}$ is the rate of growth of the momentum covariance per unit physical path length in the medium. In the following, we further clarify the connection between this tensor and the physical interpretation of its components.

\subsection{Recovering transverse momentum broadening}

To recover the usual notion of transverse momentum broadening from \Eq{Eq:qhat_tensor}, we project the covariance onto specific directions. For that purpose, let $V^\mu$ be any four-vector satisfying
\begin{equation}
V \cdot u = 0,
\end{equation}
so that $V^\mu$ is purely spatial in the fluid rest frame. We then consider the random variable
\begin{equation}
X(\lambda;V) \equiv V_\mu \left( K^\mu(\lambda) - \langle K^\mu(\lambda) \rangle \right)
= V_\mu \Delta K^\mu(\lambda)\, ,
\end{equation}
which represents the fluctuation of the jet momentum along the direction $V^\mu$. Its variance is
\begin{equation}
\label{eq:var_X}
\mathrm{Var}[X(\lambda;V)] \equiv
\langle X(\lambda;V)^2 \rangle
= V_\mu V_\nu\,\langle \Delta K^\mu \Delta K^\nu \rangle(\lambda)\, .
\end{equation}
Differentiating and using \Eq{eq:cov_evol}, we obtain (in the LRF, for a narrow high-energy jet)
\begin{equation}
\frac{d}{d\ell}\,
\mathrm{Var}[X(\ell;V)]
\simeq
V_\mu V_\nu \, \hat q^{\mu\nu}(K,u)\, .
\label{eq:var_X_deriv}
\end{equation}
Equation~\eqref{eq:var_X_deriv} encapsulates the central physical result: for any direction $V^\mu$ orthogonal to the local flow, the quadratic form $V_\mu V_\nu \hat q^{\mu\nu}$ gives, to a good approximation for a narrow, high-energy jet, the rate at which the variance of the jet momentum component along $V^\mu$ grows per unit path length in the medium.

To connect with the standard transverse broadening picture, consider the fluid rest frame and choose spatial coordinates such that the jet momentum points along the $z$-axis. Then, any purely spatial direction $V^\mu=(0,\mathbf{n})$, with $\mathbf{n}$ a unit vector, can be written as
\begin{equation}
\mathbf{n} = \cos\theta \, \hat{\mathbf e}_z + \sin\theta \, \hat{\mathbf e}_\perp,
\end{equation}
where $\hat{\mathbf e}_\perp$ is a unit vector in the transverse plane. The fluctuation of the spatial momentum along $\mathbf{n}$ is
\begin{equation}
X=\mathbf{n}\cdot\left(\mathbf{k}-\langle \mathbf{k}\rangle\right)\, ,
\label{eq:fluct}
\end{equation}
and \Eq{eq:var_X_deriv} becomes, in spatial components,
\begin{equation}
\frac{d}{d\ell}\,\langle X^2 \rangle
\simeq
n_i n_j \, \hat q^{ij}(K,u),
\label{eq:spatial}
\end{equation}
where $i,j=1,2,3$ label spatial indices in the LRF. If we choose $\mathbf{n}$ transverse to the jet direction ($\cos\theta=0$), then $n_i n_j \hat q^{ij}$ measures the transverse momentum broadening rate along that specific transverse direction. Averaging over transverse directions in an isotropic medium recovers a single scalar, the standard $\hat q$.

This analysis demonstrates that the tensor $\hat q^{\mu\nu}$ directly encodes how the jet momentum diffuses along different directions in momentum space. Its Lorentz-tensor nature allows jet broadening to be formulated covariantly, with the directional projections $V_\mu V_\nu \hat q^{\mu\nu}$ playing the role of direction-dependent generalizations of the usual $\hat q$.

\subsection{Energy broadening and $\hat q^{00}$}

Next, we turn our focus to $\hat q^{00}$ and emphasize that the tensor $\hat q^{\mu\nu}$ also encodes the \emph{energy} broadening of the jet. For studies of energy loss through elastic scattering, see, e.g.,~\cite{Peigne:2008nd,Cao:2013ita}. We define the jet energy fluctuation in the local fluid rest frame as
\begin{equation}
\Delta E_k(\lambda) \equiv E_k(\lambda) - \langle E_k(\lambda) \rangle\,.
\label{eq:energ_fluct}
\end{equation}
It follows that the energy fluctuation can be written as
\begin{equation}
\Delta E_k(\lambda) =  u_\mu \Delta K^\mu(\lambda),
\end{equation}
and its variance is
\begin{equation}
\mathrm{Var}[\Delta E_k(\lambda)]
= \left\langle \left(\Delta E_k(\lambda)\right)^2 \right\rangle
= u_\mu u_\nu
\langle \Delta K^\mu \Delta K^\nu \rangle (\lambda).
\end{equation}
Using the covariance evolution \Eq{eq:cov_evol}, we find
\begin{equation}
\frac{d}{d\ell}
\mathrm{Var}[\Delta E_k(\ell)]
\simeq
u_\mu u_\nu \, \hat q^{\mu\nu}(K,u).
\label{eq:var}
\end{equation}
In the fluid rest frame, one finds
\begin{equation}
u_\mu u_\nu \, \hat q^{\mu\nu}(K,u) = \hat q^{00}(K,u),
\end{equation}
and \Eq{eq:var} becomes
\begin{equation}
\frac{d}{d\ell}
\mathrm{Var}[\Delta E_k(\ell)] \simeq \hat q^{00}(K,u), \qquad \text{(LRF)}.
\end{equation}
Thus, $\hat q^{00}$ measures the rate at which the \emph{energy} of the jet diffuses---that is, the rate at which its energy variance grows---per unit path length in the medium. In a static, isotropic medium in its rest frame, symmetry and kinematics typically imply that $\hat q^{00}$ is suppressed for a highly energetic jet, so that transverse momentum broadening constitutes the dominant effect\footnote{
In the rest frame of a static, isotropic medium, soft elastic scatterings of a highly energetic jet satisfy $|\delta E_p| \ll |\delta \mathbf{p}_\perp|$ by kinematics. For small-angle scattering one has schematically $\delta E_p \sim q^0 \sim q_\perp^{\,2}/(2E_p)$, so that
$B^{00} \sim \langle (\delta E_p)^2 \rangle \sim \langle q_\perp^{\,4} \rangle / E_p^{\,2}$,
while $B^{ij}_\perp \sim \langle q_\perp^{\,i} q_\perp^{\,j} \rangle$.
}. However, in a flowing or out-of-equilibrium plasma, $\hat q^{00}$ need not vanish and can encode genuine energy diffusion of the jet induced by elastic scatterings with the medium. This information is lost if one only tracks the scalar transverse broadening parameter $\hat q$, which highlights an important new aspect captured by our new tensorial framework.

\subsection{Energy--momentum correlations and $\hat q^{0i}$}

We next examine the mixed components $\hat q^{0i}$, which describe correlations between energy and momentum broadening. Such mixed correlations also appear in stochastic descriptions of parton propagation, such as Langevin approaches to heavy-quark transport, where momentum and energy diffusion (and their correlations) can be discussed within a unified framework~\cite{Moore:2004tg,Peigne:2008wu}. Here, we provide a systematic covariant formulation of these ideas for highly energetic partons propagating through a medium.

To make this precise, we consider the joint fluctuations of the jet energy and of the momentum along a spatial direction $V^\mu$ orthogonal to $u^\mu$. As before, take $V^\mu$ such that $V\cdot u = 0$ and define
\begin{equation}
\Delta E_k(\lambda) =  u_\mu \Delta K^\mu(\lambda),
\qquad
X(\lambda; V) = V_\mu \Delta K^\mu(\lambda).
\end{equation}
Their covariance is
\begin{equation}
\mathrm{Cov}\!\left[\Delta E_k(\lambda), X(\lambda; V)\right]
\equiv
\left\langle \Delta E_k(\lambda)\, X(\lambda; V) \right\rangle
=
u_\mu V_\nu\,
\langle \Delta K^\mu \Delta K^\nu \rangle (\lambda).
\end{equation}
Using the covariance evolution \Eq{eq:cov_evol}, we obtain
\begin{equation}
\frac{d}{d\ell}
\mathrm{Cov}\!\left[\Delta E_k(\ell), X(\ell; V)\right]
\simeq
u_\mu V_\nu \, \hat q^{\mu\nu}(K,u).
\label{eq:cov}
\end{equation}

In the fluid rest frame, choose a purely spatial direction $V^\mu=(0,\mathbf{n})$. Then, Equation~\eqref{eq:cov} reduces to
\begin{equation}
\frac{d}{d\ell}
\mathrm{Cov}\!\left[\Delta E_k(\ell), X(\ell; \mathbf n)\right]
\simeq
n_i \, \hat q^{0i}(K,u),
\qquad \text{(LRF)}.
\end{equation}
This shows that the mixed components $\hat q^{0i}$ quantify how fluctuations in the jet energy and fluctuations in the jet momentum along direction $\mathbf n$ become correlated (or anticorrelated) as the jet propagates through the medium. In particular, a nonzero $\hat q^{0i}$ signals that kicks which deflect the jet in a given spatial direction also tend, on average, to increase or decrease its energy.

Moreover, even in a static, isotropic equilibrium medium in its rest frame, the jet direction provides a preferred axis, so symmetry alone does not force $\hat q^{0i}$ to vanish; in general one may have $\hat q^{0i}\propto \hat k^i$. For a highly energetic jet undergoing soft, small-angle elastic scatterings, however, kinematics typically suppresses energy--momentum correlations (schematically $q^0\sim q_\perp^{\,2}/(2E)$), so that $\hat q^{0i}$ is subleading at large jet energy in the strict eikonal limit. In realistic heavy-ion collisions, spacetime-dependent flow and momentum-space anisotropies can enhance and further structure $\hat q^{0i}$, encoding how energy loss and angular deflection can be dynamically tied together. Once more, such energy--momentum correlations would be entirely missed if one were to restrict the description to a single scalar jet-quenching parameter $\hat q$, illustrating a key advantage of the covariant framework introduced in this work.

\subsection{Positivity of the broadening tensor $\hat q^{\mu\nu}$ and diffusion constraints}

Since our interest in this paper is to understand the broadening tensor, we next turn to a general property that will be important for future work, namely the positivity constraints implied by a Markovian kinetic description. To this end, we consider the quadratic form built from $\hat q^{\mu\nu}(K,u)$,
\begin{equation}
Q_{\hat q}[Z] \equiv Z_\mu Z_\nu \hat q^{\mu\nu}(K,u)
= \frac{1}{E_k(K,u)}\, Z_\mu Z_\nu B^{\mu\nu}(K)
= \frac{1}{E_k(K,u)}\, Q_B[Z]\, ,
\label{eq:Qq_def}
\end{equation}
where $Z_\mu$ is an arbitrary real four-vector. Using the definition
\begin{equation}
B^{\mu\nu}(K) = \int d^4 q \, q^\mu q^\nu \, W(K,q),
\qquad W(K,q) \ge 0 ,
\label{eq:B_def}
\end{equation}
we obtain
\begin{equation}
Q_B[Z] = \int d^4 q \, Z_\mu Z_\nu \, q^\mu q^\nu \, W(K,q)
= \int d^4 q \, (Z_{\mu} q^{\mu})^2 \, W(K,q).
\label{eq:QB_sq}
\end{equation}
Here $(Z \cdot q)$ is an ordinary real number, independent of whether $Z^\mu$ or $q^\mu$ are timelike, spacelike, or null. Consequently, $(Z \cdot q)^2 \ge 0$ for all $q$, and since $W(K,q)\ge 0$ by definition, the integrand in Eq.~\eqref{eq:QB_sq} is everywhere nonnegative. Therefore,
\begin{equation}
Q_B[Z] \ge 0
\qquad
\text{for all real four-vectors } Z^\mu .
\label{eq:QB_pos}
\end{equation}

This shows that $B^{\mu\nu}(K)$ defines a \emph{positive semidefinite} symmetric bilinear form in the sense that
\begin{equation}
Z_\mu Z_\nu B^{\mu\nu}(K) \ge 0
\qquad
\text{for all } Z^\mu \in \mathbb{R}^4 .
\label{eq:B_psd}
\end{equation}
The equality $Q_B[Z]=0$ holds if and only if $Z \cdot q = 0$ for all $q$ in the support of $W(K,q)$, i.e.\ if the microscopic scatterings do not change the component of the momentum transfer along $Z^\mu$. This is a purely algebraic statement and does not depend on the Lorentzian signature of the spacetime metric.

Turning back to the covariant broadening tensor,
\begin{equation}
\hat q^{\mu\nu}(K,u) = \frac{1}{E_k(K,u)}\, B^{\mu\nu}(K),
\label{eq:qhat_def}
\end{equation}
and noting that $E_k(K,u)>0$ for a physical jet, it follows immediately that
\begin{equation}
Q_{\hat q}[Z] \ge 0
\qquad
\text{for all real four-vectors } Z^\mu ,
\label{eq:Qq_pos}
\end{equation}
i.e.\ $\hat q^{\mu\nu}(K,u)$ defines a symmetric, positive semidefinite bilinear form:
\begin{equation}
Z_\mu Z_\nu \hat q^{\mu\nu}(K,u) \ge 0
\qquad
\text{for all } Z^\mu \in \mathbb{R}^4 ,
\label{eq:qhat_psd}
\end{equation}
with equality only if the component of the momentum transfer along $Z^\mu$ vanishes for all scatterings contributing to $W(K,q)$.

In the fluid rest frame,  we may decompose $\hat q^{\mu\nu}$
into temporal and spatial parts:
\begin{equation}
\hat q^{\mu\nu} =
\begin{pmatrix}
\hat q^{00} & \hat q^{0j} \\
\hat q^{i0} & \hat q^{ij}
\end{pmatrix},
\qquad i,j = 1,2,3 .
\label{eq:qhat_decomp}
\end{equation}
Choosing $Z^\mu=(0,\mathbf V)$ with an arbitrary spatial vector $\mathbf V \in \mathbb{R}^3$, we obtain
\begin{equation}
Q_{\hat q}[Z] = V_i V_j \hat q^{ij}(K,u) \ge 0
\qquad
\text{for all } \mathbf V \in \mathbb{R}^3 .
\label{eq:qhat_spatial}
\end{equation}
Thus, the $3\times 3$ spatial matrix with entries $\hat q^{ij}$ in the LRF is positive semidefinite in the usual sense, and all of its eigenvalues are nonnegative.

This positivity condition carries physical implications that extend beyond the properties of the diffusion tensor itself. To see this, recall from \Eq{eq:cov_evol} that, in the local rest frame and for a narrow high-energy jet, the covariance evolves as
\begin{equation}
\frac{d}{d\ell}
\left\langle \Delta K^\mu \Delta K^\nu \right\rangle
\simeq
\hat q^{\mu\nu}(K,u),
\qquad
\text{LRF, narrow high-energy jet}.
\label{eq:covariance_evolution}
\end{equation}
In particular, for a purely spatial direction $V^\mu=(0,\mathbf n)$ we have
\begin{equation}
\frac{d}{d\ell}
\operatorname{Var}\!\left[X(\ell;\mathbf n)\right]
=
\frac{d}{d\ell}
\left\langle X(\ell;\mathbf n)^2 \right\rangle
\simeq
n_i n_j\, \hat q^{ij}(K,u),
\qquad
X(\ell;\mathbf n)
=
\mathbf n \cdot \left( \mathbf{k}(\ell) - \langle \mathbf{k}(\ell) \rangle \right),
\label{eq:directional_variance}
\end{equation}
where $\ell$ is the path length in the LRF and $\mathbf n$ is a unit vector. The positivity condition then implies
\begin{equation}
\frac{d}{d\ell}
\operatorname{Var}\!\left[X(\ell;\mathbf n)\right]
\ge 0
\qquad
\text{for all spatial directions } \mathbf n .
\label{eq:variance_positivity}
\end{equation}
In other words, within the Markovian kinetic description encoded by the positive kernel $W(K,q)$:
\begin{enumerate}[label=(\roman*)]
\item The variance of the jet momentum along any direction in momentum space cannot \emph{decrease} due to the stochastic scatterings. In fact, scatterings can only broaden the distribution (or leave a given component unaffected), but cannot ``sharpen'' it.
\item The eigenvalues of $\hat q^{ij}$ in the LRF are the principal momentum-space diffusion coefficients, one for each principal axis of broadening, and their nonnegativity expresses the fact that diffusion increases fluctuations along every principal direction.
\end{enumerate}

These properties can be viewed as \emph{diffusion-type constraints} imposed by the underlying kinetic theory on any phenomenological model of jet broadening. If one attempts to extract an effective $\hat q^{\mu\nu}$ from data or from a macroscopic model (e.g.\ hydrodynamics plus some ansatz for jet--medium coupling), then requiring that $\hat q^{\mu\nu}$ be realizable as the second moment of a positive rate $W(K,q)$ implies that $\hat q^{\mu\nu}$ must be symmetric and positive semidefinite. In particular, in the LRF the spatial matrix $\hat q^{ij}$ must have nonnegative eigenvalues. Any negative eigenvalue (or negative quadratic form $V_i V_j \hat q^{ij}$ for some $\mathbf V$) would signal a breakdown of the simple Markovian kinetic picture in terms of real scatterings, pointing either to strong quantum coherence effects (beyond a Boltzmann/Fokker--Planck description) or to an inconsistency in the modeling.

Thus, just as classical energy conditions constrain the allowed forms of the stress--energy tensor in a macroscopic effective theory~\cite{HawkingEllis}, the positivity of $\hat q^{\mu\nu}$ provides \emph{diffusion conditions} that any physically realizable jet-broadening tensor in kinetic theory must satisfy. We will not pursue a detailed analysis of these constraints in the present work, and instead focus on applying the formalism developed above to a simple toy model. In particular, we study the jet momentum broadening tensor in a $\lambda\varphi^4$ theory and use it to investigate how non-equilibrium effects manifest themselves in two regimes: quantum statistics and the classical limit. This allows us to make concrete physical statements about the system while leaving a more detailed exploration of the broader consequences of the positivity condition to future work.

\subsection{Summary of physical content of $\hat q^{\mu\nu}$}

The Lorentz-covariant definition of jet momentum broadening, $\hat q^{\mu\nu}$, encodes the rate of growth of the jet momentum covariance per unit physical path length in the medium, as discussed in Sec.~\ref{Sec:kinetic_theory}. Unlike the scalar jet-quenching parameter, this tensorial formulation captures additional information about how energy and momentum are exchanged between the jet and the medium. For clarity, we summarize the physical interpretation of the main components discussed in this section:

\begin{enumerate}[label=(\roman*)]
\item $\hat q^{00}$ measures the rate at which the \emph{energy} of the jet diffuses, i.e.\ the growth rate of its energy variance per unit path length. For a highly energetic jet undergoing soft, small-angle elastic scatterings in a static, isotropic medium in its rest frame, $\hat q^{00}$ is kinematically suppressed, so that transverse momentum broadening constitutes the dominant effect.

\item $\hat q^{0i}$ measures energy--momentum correlations, i.e.\ how fluctuations in the jet energy and fluctuations in the jet momentum along a spatial direction are correlated (or anticorrelated) as the jet propagates through the medium.

\item $\hat q^{ij}$ measures the rate of momentum broadening in the spatial sector. In particular, for a unit vector $\mathbf n$ in the LRF, the quadratic form $n_i n_j\,\hat q^{ij}$ gives the broadening rate along that direction, and averaging over transverse directions in an isotropic medium recovers a single scalar, the standard $\hat q$.
\end{enumerate}

\medskip
Finally, we emphasize that the spatial components $\hat q^{ij}$ form a positive semidefinite matrix in the local rest frame, with all eigenvalues nonnegative. This implies general diffusion-type constraints on kinetic-theory descriptions of jet--medium interactions and provides insight into the regime of applicability of the diffusion-based formalism, particularly in out-of-equilibrium settings.

\section{Covariant jet momentum broadening in \(\lambda\varphi^4\) theory }
\label{Sec:quant_qhat}

Having established a general, covariant framework for jet momentum broadening and clarified the physical meaning of each component of $\hat{q}^{\mu\nu}$ in Sec.~\ref{Sec: Fokker_Planck}, we
now apply this formalism to an explicit microscopic model, (tree-level) massless scalar theory with $\lambda\varphi^4$ quartic interactions. The goal of this section is twofold: first, to demonstrate how the tensorial structure derived in Secs.~\ref{Sec:kinetic_theory}--\ref{Sec: Fokker_Planck} emerges in a concrete calculation; and second, to identify which components of $\hat{q}^{\mu\nu}$ are most sensitive to non-equilibrium effects.

For that, we work with bosonic particles, which means that we take the statistical factor in $(1+a\,f)$ to be $a=1$ in \Eq{eq:loss_term}. In that sense, this toy-model provides some insight into more complex systems, and has the advantage of allowing for analytical calculations, and additionally for a simple tensorial description to be made, such that we are able to compute this tensor explicitly for the first time. 

\begin{figure}
    \centering
    \includegraphics[width=0.4\linewidth]{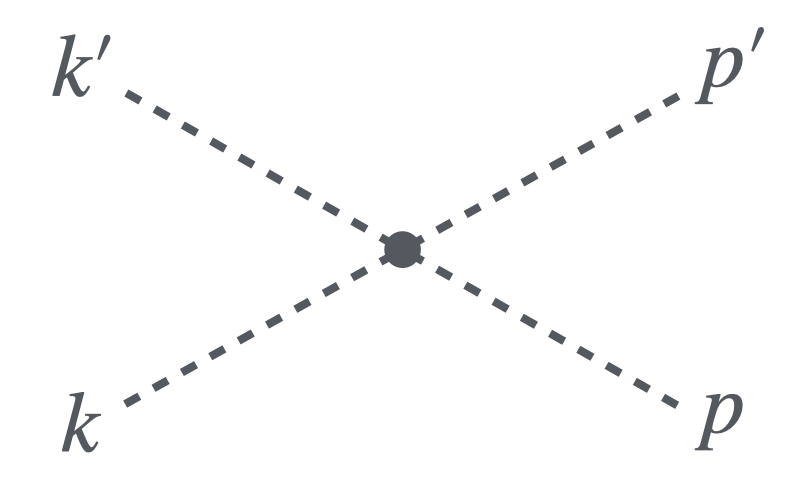}
    \caption{Schematic representation of the $\lambda\varphi^4$ elastic scattering.}
    \label{fig:phi4_interaction}
\end{figure}

Since the functional form of the differential cross section depends on the specifics of the interaction, we adopt the same form employed in~\cite{Denicol:2022bsq,Rocha:2024cge}. To this end, we consider a system of massless scalar particles governed by the Lagrangian density:
\begin{equation}
\label{eq:lag-phi4}
\mathcal{L} = \frac{1}{2} \partial_{\mu} \varphi \ \partial^{\mu} \varphi
-
\frac{\lambda \varphi^{4}}{4!}.
\end{equation}
The interaction is shown schematically in Fig.~\ref{fig:phi4_interaction}. At leading order in the coupling constant, this interaction is simply given by~\cite{Peskin:1995ev}, 
\begin{equation}
\label{eq:cross-sec-phi4}
\begin{aligned}
& \sigma_{T}(s) = \frac{1}{2} \int d\Phi\, d\Theta \, \sin\Theta \, \sigma(s, \Theta) = \frac{\lambda^2}{32 \pi s} \equiv \frac{g}{s}\, ,
\end{aligned}    
\end{equation}
where $\Phi$ is the azimuthal angle in the center-of-mass frame and $g \equiv \lambda^{2}/(32 \pi)$, and $s$ is again the Mandelstam variable, for which we use the same definition as in Section~\ref{Sec:kinetic_theory}. Hence, the transition rate, as a function of $\sigma_T(s)$, is given by, 
\begin{align}
     \mathcal{W}_{a} =(2\pi)^6\delta^{(4)}(K+K'-P-P')\sigma_T(s)\, s\, .
     \label{Eq:lambda_phi4_w}
\end{align}

We begin by introducing the key definitions used. Here, $s$ is the Mandelstam variable $s \equiv (K^{\mu}+K'^{\mu})(K_{\mu}+K'_{\mu}) = (P^{\mu}+P'^{\mu})(P_{\mu}+P'_{\mu})$, which denotes the total center-of-mass energy squared.
To carry out this calculation, we recover the expression for the loss term from \Eq{eq:loss_term} in a more explicit formulation, including already the Bose–Einstein statistics. This leads to,
\begin{equation}
    \begin{aligned}
        \mathcal{C}_{loss}[f_{\mathbf{k}}] &=  \frac{1}{2}\int dP \, dK' \, dP' \, \sigma_T(s)\, s\, (2\pi)^4\delta^{(4)}(K+K'-P-P') \, f_{{\mathbf{k}}}f_{{\mathbf{k'}}}(1+f_{{\mathbf{p}}'})(1+f_{{\mathbf{p}}})\, .
    \end{aligned}
\end{equation}
Next, we follow the procedure described in Section~\ref{Sec:kinetic_theory} to reproduce the jet quenching tensor. To that end, we assume the momentum of the jet is much larger than the average thermal momentum,  $|\mathbf{k}|\gg \langle p'\rangle$, such that we can take \( f_{\mathbf{p}} \approx 0 \), similarly to what was done in~\cite{Boguslavski:2023waw}, and we consider only isotropic distributions throughout these calculations. Replacing \Eq{Eq:lambda_phi4_w}, the loss term can be explicitly written as:
\begin{equation}
\begin{aligned}
    \hat{q}^{\mu\nu}(K^\mu)=\frac{g}{2 \, E_{\mathbf{k}}}\int dP \, dK' \, dP'q^\mu q^\nu\delta^{(4)}(K+K'-P-P')\, f_{\mathbf{k'}}[1+f_{\mathbf{p'}}]\, .
    \end{aligned}
\end{equation}
Next, for the sake of clarity, we begin by breaking this integral into four pieces using the product, and introduce the notation that will be used for these calculations,
\begin{equation}
\begin{aligned}
  q^\mu q^\nu &= (K - P)^\mu (K - P)^\nu \\
  &= \underbrace{K^\mu K^\nu}_{\text{(i)}} 
     - \underbrace{K^\mu P^\nu}_{\text{(ii)}} 
     - \underbrace{K^\nu P^\mu}_{\text{(iii)}} 
     + \underbrace{P^\mu P^\nu}_{\text{(iv)}} \, .
\end{aligned}
\label{Eq:exchanged_momentum}
\end{equation}
Within $\lambda\varphi^4$ at leading order, $\sigma_T(s)$ yields finite elastic integrals without additional IR regulators, such that we impose no constraint on the momentum exchanged between the jet and the medium. For each term, we perform the integrals for a generic form of the momenta of each parton involved, leaving only one input, which is the form of the jet four-momentum, that we assume to be large in comparison with the medium partons, but a fixed finite $k$. We begin with term $(i)$, which is computed making use of the fact that these calculations involve only massless particles, meaning $E_{\mathbf{k}}=|\mathbf{k}|$, such that the integral for $(i)$ can be written as,
\begin{equation}
    \begin{aligned}
        (i)&= \frac{g}{2\,E_{\mathbf{k}}}\int dP \, dK' \, dP' K^\mu K^\nu (2\pi)^5\delta^{(4)}(K+K'-P-P')\, f_{\mathbf{k'}}[1+f_{\mathbf{p'}}]\, ,
    \end{aligned}
\end{equation}
The integral in $\mathbf{p}$ can be directly performed using the delta function, yielding 
\begin{equation}
    \begin{aligned}
        (i)&= \frac{g}{2\,E_{\mathbf{k}}}\int  \, dK' \, dP' K^\mu K^\nu (2\pi)^2\delta(\sqrt{s}-p_0-p'_0)\, f_{\mathbf{k'}}[1+f_{\mathbf{p'}}]\, ,
    \end{aligned}
\end{equation}
Since we consider isotropic distributions, the integration over \( \mathbf{p}' \) is performed using the energy delta function, and it yields,
\begin{equation}
    \begin{aligned}
        (i)&= \frac{g}{2\,E_{\mathbf{k}}}\int  \, dK' \, K^\mu K^\nu\, f_{\mathbf{k'}}[1+f_{\mathbf{k}+\mathbf{k'}}]\, .
    \end{aligned}
\end{equation}
As mentioned above, we assume that \( k \gg k' \). This ensures that the integration of \( k' \) is naturally regulated, since the limit \( k \rightarrow 0 \) corresponds to the absence of a jet, and \( k \) is taken to have a fixed value throughout this calculation. Therefore, we can simply write the final expression as a function of the integration over the distribution of particles in the medium as, 
\begin{equation}
    \begin{aligned}
    \label{Eq:temr1_glue}
        (i) &= \frac{g K^\mu K^\nu}{E_{\mathbf{k}}} \left( \int  \, dK' \,  f_{\mathbf{k}'} + \int  \, dK' \,  f_{\mathbf{k}'} f_{\mathbf{k} + \mathbf{k}'} \right) \, .
    \end{aligned}
\end{equation}
where we can use an analogous definition of the in-medium effective mass of the scalar as given in Ref.~\cite{Kurkela:2018oqw} to interpret the result obtained in the expression above. We also note that this term represents the first non-equilibrium contribution in our calculations, and the tensorial structure that was described in Sec.~\ref{Sec: Fokker_Planck}. In order to connect these two, we define an effective mass $m_{\varphi}$ using:
\begin{equation}
    \begin{aligned}
        m_{\varphi}^2 = \int  \, dK' \,f_{\mathbf{k}'} \, .
        \label{eq:varphi_mass}
    \end{aligned}
\end{equation}
We emphasize that $m_\varphi$ is an effective screening–like, in-medium scale that naturally emerges in our calculations, as commonly discussed in finite-temperature field theory~\cite{Kapusta:2006pm}. Additionally, the coupling dependence resides in $g=\lambda^2/(32\pi)$ and has been kept out of this definition to simplify the notation.
With that, we rewrite expression~\Eq{Eq:temr1_glue} as,
\begin{equation}
    \begin{aligned}
       (i) &= \frac{ gK^\mu K^\nu}{E_{\mathbf{k}}} m_{\varphi}^2 + \frac{g K^\mu K^\nu}{E_{\mathbf{k}}} \int  \, dK' \,  f_{\mathbf{k}'} f_{\mathbf{k} + \mathbf{k}'} \, .
    \end{aligned}
\end{equation}
Additionally, we note that the second term vanishes in the limit of \( k\gg T \). This can be simply observed by noting that the biggest contribution to the integration is given at small values of $k'\rightarrow 0$, such that, once a fixed value of $k$ is added, this contribution becomes arbitrarily small. Therefore, it is straightforward to write an approximate final form for term~(i) as,
\begin{equation}
    \begin{aligned}
        (i) & \approx \frac{ gK^\mu K^\nu}{E_{\mathbf{k}}} m_{\varphi}^2\, .
    \end{aligned}
\end{equation}
We note that the in-medium mass here includes the effects of non-equilibrium interactions, meaning that the mass can change whenever an out-of-equilibrium distribution function is included in this calculation. Therefore, terms that contain an in-medium mass will carry the information of any effects that are included for the interaction between the jet and the medium. We now repeat the procedure for the remaining terms. Next, we compute term~$(ii)$:
\begin{widetext}
\begin{align}
(ii)= \frac{g}{2\,E_{\mathbf{k}}}K^\mu \int  \, dK' \,f_{\mathbf{k}'}\int  \, dP \, dP'\, \delta^{(4)}\!\left(P_T - P - P'\right)\, P^\nu\, [1+f_{\mathbf{p}'}]\, ,
\label{II}
\end{align}
\end{widetext}
where $P_T^\mu=\sqrt{s}\,t^\mu$, $t^\mu \equiv (1,\vec{0}\,)$ and we have used the properties of the delta function to separate the integral over $\mathbf{p}$ and $\mathbf{p}'$, which now only depends on $P^\mu$ and $P'^\mu$. This can be easily computed as follows,
\begin{align}
    \int  \, dP \, dP' \,\delta^{(4)}\!\left(P_T - P - P'\right) P^\nu \left[ 1 + f_{\mathbf{p}'} \right] &=\int  \, dP \, dP' \, \frac{(2\pi)^5}{(2\pi)^6}\delta(\sqrt{s} -p -p')\,\delta^{(3)}(\mathbf{p}+\mathbf{p}')\, P^\nu \, [1+f_{\mathbf{p}'}]\\
    &=2\sqrt{s}\, t^\nu \, [1+f_{\sqrt{s}}]\, .
\end{align}
Here, we use the definition of energy in the center of mass to write $s=(K+K')^\mu(K+K')_\mu$ and perform the remaining integral, defined as,
\begin{align}
    (ii)&=\frac{g \, K^\mu  }{\,E_{\mathbf{k}}}\int \, dK' \,f_{\mathbf{k}'}[1+f_{\mathbf{k}+\mathbf{k}'}] (K+K')^\nu\, ,
\end{align}
where the calculation is analogous to term~$(i)$, and it leads to,
\begin{align}
    (ii)&= \frac{g }{\,E_{\mathbf{k}}}\left(K^\mu K^\nu m_{\varphi}^2 + K^\mu J^\nu + K^\mu K^\nu \int \, dK' \,  f_{\mathbf{k}'} f_{\mathbf{k} + \mathbf{k}'} + K^\mu\int  \,dK' \, K'^\nu f_{\mathbf{k}'} f_{\mathbf{k} + \mathbf{k}'}\right)\, .
\end{align}
where $J^\nu$ is defined as the particle number diffusion current, given by,
\begin{equation}
\begin{aligned} 
\label{Eq:diff}
J^\nu = \int dK\, K^\nu f_{\mathbf{k}}\, ,
\end{aligned}    
\end{equation}
We note that, for isotropic distributions in the local rest frame, the particle-number current has the fixed form $J^\nu=nu^\nu$, such that, non-equilibrium corrections may change the number of particles, $n$, but will not affect the tensorial structure of the current $J^\nu$. Therefore, the final form of term~$(ii)$ leads to an interesting conclusion: although \( J^\nu \) is not affected by the out-of-equilibrium nature of the medium in the case of isotropic distributions, there is still a contribution from non-equilibrium effects through the in-medium mass. Additionally, we obtain an extra term that carries a contribution similar to the one observed for term~$(i)$, which is infrared regulated and small, due to the large jet momentum $\mathbf{k}$ in the distribution function, as discussed earlier in this section.

Next, we compute term~$(iii)$ by simply noting the symmetry between terms~$(ii)$ and~$(iii)$, related by \( \mu \leftrightarrow \nu \). Therefore, we can directly write the result for term~$(iii)$ as:
\begin{align}
    (iii)&= \frac{g }{\,E_{\mathbf{k}}}\left(K^\mu K^\nu m_{\varphi}^2 + K^\nu J^\mu + K^\mu K^\nu \int  \, dK' \,  f_{\mathbf{k}'} f_{\mathbf{k} + \mathbf{k}'} + K^\mu \int \, dK' \,  K'^\nu f_{\mathbf{k}'} f_{\mathbf{k} + \mathbf{k}'}\right)\, .
\end{align}
We now turn to term~$(iv)$, which is more involved but follows the same integration type as before. For completeness, we reproduce the expression for term~$(iv)$ here,
\begin{equation}
    \begin{aligned}
    \label{Eq:iv}
    (iv)=\frac{g }{2\,E_{\mathbf{k}}}\int  \, dK' \,f_{\mathbf{k}'} \int  \, dP \, dP' \,\delta^{(4)}\!\left(P_T - P - P'\right)\, P^\mu P^\nu \, [1+f_{\mathbf{p}'}]\, ,
\end{aligned}
\end{equation}
where we can use the definition $\Delta^{\mu\nu}_T=g^{\mu\nu}-\frac{P_T^\mu P_T^\nu}{s}$ to rewrite this integral in terms of a scalar integral and a known tensor as,
\begin{equation}
    \begin{aligned}
       A^{\mu\nu}=\int  \, dP \, dP' \,(2\pi)^5\delta^{(4)}\!\left(\sqrt{s}\, t^\mu -P^\mu -P'^\mu\right) P^\mu P^\nu \, [1+f_{\mathbf{p}'}] = 2A\, [1+f_{P_T}] \left(\frac{P_T^\mu P_T^\nu}{s}-\frac{\Delta^{\mu\nu}_T}{3}\right)\, ,
       \label{eq:trick}
    \end{aligned}
\end{equation} 
This gives rise to a term that is directly dependent on the energy–momentum tensor defined as, 
\begin{equation}
\begin{aligned}
\label{Eq:energ_tensor}
T^{\mu \nu} = \int dK\, K^{\mu} K^{\nu} f_{\mathbf{k}}\, .
\end{aligned}    
\end{equation}
We note that, again, for isotropic distributions, this tensor will not be affected by any non-equilibrium corrections (using the choice of matching conditions made in \cite{Bazow:2015dha,Mullins:2022fbx}). The clear separation between equilibrium contributions and non-equilibrium corrections is an advantage that arises as a consequence of the tensorial description.
To summarize, we write the expression for \(\hat{q}^{\mu\nu}\) as a function of the hydrodynamical variables of interest, and the in-medium mass,
\begin{equation}
    \begin{aligned}
    \hat{q}^{\mu\nu}(K^\mu) &= \frac{g }{E_{\mathbf{k}}}\Bigg[-\frac{2}{3}\left(K^\mu J^\nu +K^\nu J^\mu \right)+\frac{2}{3}T^{\mu\nu} - \frac{K^\alpha J_\alpha}{3} g^{\mu\nu} +\frac{1}{6}K^\mu K^\nu m_{\varphi}^2 \\
    &\hspace{2.5em}-\frac{2}{3}K^\mu\left(K^\nu  \int\, dK'\,  f_{\mathbf{k}'} f_{\mathbf{k} + \mathbf{k}'} + \int\, dK'\,  K'^\mu f_{\mathbf{k}'} f_{\mathbf{k} + \mathbf{k}'}\right)+\frac{2}{3} \int\, dK'\,  K'^\mu K'^\nu f_{\mathbf{k}'} f_{\mathbf{k} + \mathbf{k}'} + \cdots \Bigg]\, .
\end{aligned}
\end{equation}
For simplicity, we choose to focus on the leading contributions, therefore not explicitly writing the contributions from the terms containing an integration over $f_{\mathbf{k} + \mathbf{k}'}$, since, as explained before, these terms are small enough to be discarded in the high-momentum jet limit. For sufficiently large values of $k$, $\hat{q}^{\mu\nu}$ takes the simplified form:
\begin{equation}
\label{Eq:qhat_glue}
\hat{q}^{\mu\nu}(K^\mu) \approx \frac{g}{E_{\mathbf{k}}}\Bigg[
-\frac{2}{3}\left(K^\mu J^\nu + K^\nu J^\mu\right)
+ \frac{2}{3} T^{\mu\nu}
- \frac{\, K^\alpha J_\alpha}{3}\, g^{\mu\nu}
+ \frac{1}{6} K^\mu K^\nu\, m_\varphi^2
\Bigg] .
\end{equation}

We observe that, even through simple dimensional analysis, one can see that in a 
\(\lambda\varphi^4\) theory, the out-of-equilibrium contribution enters at leading order. As a result, there is a significant correction to the jet momentum broadening, which we will study in more detail in the following sections. For clarity, we explicitly separate the equilibrium contribution
from the non--equilibrium correction as 
\begin{equation}
\hat{q}^{\mu\nu}(K^\mu)_{\rm eq} \approx \frac{g}{E_{\mathbf{k}}}\Bigg[
-\frac{2}{3}\left(K^\mu J^\nu + K^\nu J^\mu\right)
+ \frac{2}{3} T^{\mu\nu}
- \frac{K^\alpha J_\alpha}{3}\, g^{\mu\nu}
\Bigg] ,
\end{equation}
and
\begin{equation}
\hat{q}^{\mu\nu}(K^\mu)_{\rm neq} \approx 
\frac{g}{E_{\mathbf{k}}}\frac{1}{6} K^\mu K^\nu\, m_\varphi^2 .
\end{equation}
Here, we observe the structure discussed in Sec.~\ref{Sec: Fokker_Planck}
naturally emerging from the calculations. Of course, given the simplicity of our toy model, the non--equilibrium contributions to energy broadening and energy--momentum correlations are the same as those for the traditional transverse momentum broadening; however, we emphasize that this need not be the case in more general settings.

Finally, we note that this final expression provides a simple pocket formula that can be used to describe
jet–medium interactions in a non-equilibrated system considering a $\lambda\varphi^4$ theory. Since the final expression is governed by only hydrodynamic quantities and the in-medium mass, we note that the same quantities should also appear in an approximate description of the system using classical statistics. This is discussed in detail below.

\subsection{The classical limit}
\label{Sec:phi^4}

The classical calculation follows exactly the same structure as the quantum one,
with the statistical factor $(1+a\,f)$ now reduced to unity ($a=0$).
Thus, all kinematic relations and integrals remain as in
Sec.~\ref{Sec:quant_qhat}, and only the distribution function changes to the
Boltzmann form. Here, the momentum exchanged between two scattering partons is given as in \Eq{Eq:exchanged_momentum}. For clarity, we use the same notation as in Section~\ref{Sec:quant_qhat}. 

In the classical limit, the loss term from \Eq{eq:loss_term} can be written as:
\begin{equation}
    \begin{aligned}
        \mathcal{C}_{loss}[f_{\mathbf{k}}] &=  \frac{1}{2}\int dP \, dK' \, dP' \, \mathcal{W}_{kk' \rightarrow pp'} (f_{\mathbf{k}}f_{\mathbf{k'}})\, .
    \end{aligned}
\end{equation}
Next, we replace \Eq{Eq:lambda_phi4_w} in the loss term, and follow the derivation from Section~\ref{Sec:kinetic_theory} to obtain the expression for $\hat{q}^{\mu\nu}$ as
\begin{equation}
\begin{aligned}
    \hat{q}^{\mu\nu}(K^\mu)=\frac{1}{2 \, E_{\mathbf{k}}}\int dP \, dK' \, dP' \, q^\mu q^\nu (2\pi)^5\delta^{(4)}(K+K'-P-P')\, \sigma_T(s)\, s\, f_{\mathbf{k'}}\, .
    \end{aligned}
\end{equation}
Analogously to the previous section, we divide our calculations into four terms to be computed using similar techniques, such that the first term can be written as
\begin{equation}
    \begin{aligned}
        (i)&= \frac{g}{2\,E_{\mathbf{k}}}\int dP \, dK' \, dP' \, K^\mu K^\nu (2\pi)^5\delta^{(4)}(K+K'-P-P')\, f_{\mathbf{k'}}\, ,
    \end{aligned}
\end{equation}
where we already replaced \Eq{eq:cross-sec-phi4} into the expression above, and performed the appropriate simplifications. We note that the integrations over $\mathbf{p}$ and $\mathbf{p}'$ are straightforward and can be performed using $\delta^{(4)}(K+K'-P-P')$. Here, we leave the final integration over $\mathbf{k}'$ for later and denote it by $m_{\varphi}^2$. This is interesting because it is a quantity that contains the non-equilibrium corrections for isotropic distributions and has an analogous interpretation to an in-medium effective mass from \Eq{eq:varphi_mass}. The computation of $(i)$ is straightforward, and it yields
\begin{align}
   (i)=\frac{gK^\mu K^\nu\,}{E_{\mathbf{k}}} m_{\varphi}^2 \, .
   \label{eq:(i)}
\end{align}
We note that term $(i)$ will be affected by non-equilibrium corrections to the medium, similarly to what was observed for the quantum statistics case. For the second term, the integrals are given by
\begin{widetext}
\begin{align}
(ii)= \frac{g K^\mu}{2\,E_{\mathbf{k}}}\int \, dK' \, f_{\mathbf{k}'}\int dP \, dP' \,(2\pi)^5\delta^{(4)}\!\left(P_T - P - P'\right) P^\nu\, ,
\label{II-c}
\end{align}
\end{widetext}
Since the explicit calculation was carried out in the previous section, we note that the same techniques apply to a classical distribution of particles. For consistency, we use the same definitions of the Mandelstam variables as used in Section~\ref{Sec:quant_qhat}, which leads to
\begin{align}
    (ii)&=\frac{g\, K^\mu}{\,E_{\mathbf{k}}}\int \, dK' \, f_{\mathbf{k}'} (K+K')^\nu\, ,
\end{align}
where we note that the first term of this integral is straightforward, since there is no dependence on $K^\nu$. Again, the second term can be written as a function of the diffusion current $J^\nu$, giving rise to the same equilibrium contribution observed in Sec.~\ref{Sec:quant_qhat}, and leaving all the information related to the jet--medium interaction to be carried by the in-medium effective mass as
\begin{align}
    (ii)&= \frac{gK^\mu K^\nu}{E_{\mathbf{k}}} m_{\varphi}^2 + \frac{g}{E_{\mathbf{k}}}K^\mu J^\nu \, .
\end{align}
Next, we compute term $(iii)$ by simply noting the symmetry between $(ii)$ and $(iii)$, which can be obtained by switching $\mu\leftrightarrow\nu$. Therefore, we can directly write the answer for $(iii)$ as
\begin{align}
    (iii)= \frac{gK^\nu K^\mu}{\,E_{\mathbf{k}}}  m_{\varphi}^2 + \frac{g  }{\,E_{\mathbf{k}}}K^\nu J^\mu \, ,
\end{align}
where we note that, so far, all terms are affected by non-equilibrium medium corrections, showing that this effect is not subleading and will matter for the study of jet--medium interactions, especially since the jet is expected to be created in the early stages of the collision, when the system is still far from equilibrium. The remaining term, denoted as $(iv)$, is given by
\begin{align}
    (iv)&=\frac{g}{2E_{\mathbf{k}}}\int \, dK' \, f_{\mathbf{k}'} \int dP \, dP' \,\delta^{(4)}\!\left(P_T - P - P'\right) P^\mu P^\nu\, ,
\end{align}
which can be computed using similar techniques to those employed for the first two terms. Since the procedure is similar to that for \Eq{Eq:iv}, we will not explicitly write it here.

Finally, we combine the results from all the terms computed in this section and obtain an expression for $\hat{q}^{\mu\nu}(K)$, which again depends on the Debye screening-like term and the hydrodynamic quantities $J^\mu$ and $T^{\mu\nu}$,
\begin{equation}
    \begin{aligned}
    \hat{q}^{\mu\nu}(K^\mu) &= \frac{g}{E_{\mathbf{k}}}\left[-\frac{2}{3}\left(K^\mu J^\nu +K^\nu J^\mu \right)+\frac{2}{3}T^{\mu\nu} - \frac{K^\alpha J_\alpha}{3} g^{\mu\nu} +\frac{1}{6}K^\mu K^\nu m_{\varphi}^2 \right]\, .
    \label{Eq:q_hat_lambdaphi4}
\end{aligned}
\end{equation}
We note that expression \Eq{Eq:q_hat_lambdaphi4} is completely analogous to \Eq{Eq:qhat_glue}. This shows that, for sufficiently large jet momentum in the present $\lambda\varphi^4$ setting, quantum statistical effects in the medium become subleading and the resulting broadening tensor $\hat q^{\mu\nu}$ is well approximated by its classical (Boltzmann) counterpart. In other words, in the high-momentum jet regime we find agreement between the $\hat q^{\mu\nu}$ obtained using Bose--Einstein statistics and the one obtained in the classical limit, which motivates using the classical description as a reliable and analytically tractable approximation. This type of correspondence between quantum and classical descriptions of momentum broadening has also been discussed in~\cite{Majumder:2007hx}, in a different framework, when comparing higher-twist resummation and a classical Lorentz--Langevin picture. With this motivation in mind, in the next section we use results from  \cite{Mullins:2022fbx} to study isotropic out-of-equilibrium scenarios  in which we can explicitly compute how non-equilibrium distributions affect jet quenching within the classical kinetic description.


\subsection{Non-equilibrium medium corrections}
\label{sec:outofeq}

In this section we analyze the impact of an out-of-equilibrium medium on jet momentum broadening. Specifically, we consider non-equilibrium effects in the $\lambda\varphi^4$ theory discussed above using the expression in Eq.~\eqref{Eq:q_hat_lambdaphi4}. In this analysis, we isolate terms sensitive to deviations from equilibrium, enabling a direct comparison of equilibrium and non-equilibrium contributions to jet quenching. The non-equilibrium corrections are explicitly separated in Eq.~\eqref{Eq:q_hat_lambdaphi4}, and for clarity, we denote them as $\delta\hat{q}^{\mu\nu}(K)$. These contributions are summarized in the expression below,
\begin{widetext}
    \begin{align}
    \delta\hat{q}^{\mu\nu}(K^\mu) &= \frac{g}{E_{\mathbf{k}}}\left[ \frac{1}{6}K^\mu K^\nu\int_{\mathbf{k}'}f_{\mathbf{k}'}\right]\, .
    \label{eq:delta_qhat_phi4}
\end{align}
\end{widetext}
Here, we define the four-momentum $K^\mu$ as 
\begin{equation}
\begin{aligned}
    K^\mu &= E_k\, u^\mu + K^{\langle \mu \rangle} \, ,
\end{aligned}
\label{eq:P_definition}
\end{equation} 
so that the computation of $K^\mu K^\nu$ is straightforward and yields
\begin{equation}
\begin{aligned}
    K^\mu K^\nu  &= (E_k u^\mu + K^{\langle \mu \rangle})(E_k u^\nu + K^{\langle \nu \rangle})
    = E_k^2 u^\mu u^\nu + E_k \big( u^\mu K^{\langle \nu \rangle} + u^\nu K^{\langle \mu \rangle} \big) + K^{\langle \mu \rangle}K^{\langle \nu \rangle}  \, .
\end{aligned}
\label{eq:P_expansion}
\end{equation}
Plugging this back into \Eq{eq:delta_qhat_phi4} gives
\begin{equation}
\begin{aligned}
    \delta\hat{q}^{\mu\nu}(K^\mu) &= \frac{g}{6E_{\mathbf{k}}}\left(u^\mu u^\nu E_k^2\int_{\mathbf{k}'}  f_{\mathbf{k}'} + u^\mu E_k K^{\langle \nu \rangle} \int_{\mathbf{k}'} f_{\mathbf{k}'} +  u^\nu E_k K^{\langle \mu \rangle}\int_{\mathbf{k}'} f_{\mathbf{k}'} +  K^{\langle \mu \rangle} K^{\langle \nu \rangle}\int_{\mathbf{k}'}   f_{\mathbf{k}'} \right) \, .
\end{aligned}
\label{eq:Q_tensor_expanded}
\end{equation}
At this point, we note that all non-equilibrium effects enter through the integral over $f_{\mathbf{k}'}$, while the tensor structure can be analyzed separately. Thus, one can choose a convenient reference frame to perform this analysis without loss of generality. This means that the size of the jet quenching observed is controlled by the in-medium mass of the particles involved in each collision throughout the propagation of the highly energetic parton.
Finally, this leads to a compact expression for the full correction, which can be evaluated in any frame as
\begin{equation}
\begin{aligned}
    \delta\hat{q}^{\mu\nu}(K^\mu) &= \frac{g\,m_{\varphi}^2}{6E_{\mathbf{k}}} \left( u^\mu u^\nu E_k^2 + u^\mu E_k K^{\langle \nu \rangle} + u^\nu E_k K^{\langle \mu \rangle} + K^{\langle \mu \rangle} K^{\langle \nu \rangle} \right)  \, ,
\end{aligned}
\label{eq:Q_final_phi4}
\end{equation}
where $m_{\varphi}^2$ is the in-medium mass of the scalar defined earlier.

We note that the in-medium mass can be interpreted as a screening effect. When non-equilibrium distributions are included in the calculation, a screening mechanism emerges that modifies the momentum exchange between scattering partners in the medium. This is particularly interesting, as similar screening effects arise in non-Abelian gauge theories such as QCD, affecting not only $2 \rightarrow 2$ processes that govern jet momentum broadening in heavy-ion collisions, but also splitting processes that dominate the energy loss of these high-energy partons. From this point forward, we investigate the impact of different non-equilibrium distributions on this phenomenon.

In the following, we compute the impact of a non-equilibrium distribution on our expression for $\delta\hat{q}^{\mu\nu}$ in \Eq{eq:Q_final_phi4}, using well-established analytic and numerical solutions to the Boltzmann equation from a setup similar to the one considered in this work, originally presented in \cite{Mullins:2022fbx}. In that work, the relativistic Boltzmann equation is considered in an expanding, flat FLRW spacetime \cite{Weinberg:1972kfs}, 
\begin{equation}
    K^\mu \partial_\mu f_{\mathbf{k}} + K^\mu \Gamma^\lambda_{\mu\nu} K^\nu \frac{\partial f_{\mathbf{k}}}{\partial K^\lambda} = C[f_{\mathbf{k}}] , 
\end{equation}
with tree-level $\lambda \varphi^4$ interactions. The symmetries of this setup indicate that the distribution function depends only on the momentum through $E_k$ and on time, so we write $f_{\mathbf{k}}(t)$. Then, it is convenient to write the distribution function as a series expansion in the associated Laguerre polynomials
\begin{equation}
    f_{\mathbf{k}}(t) = f_{\mathbf{k}}^{\mathrm{eq}} \sum_{n=0}^{\infty} c_n(t) L_n^{(2)}(E_k) ,
\end{equation}
where $f_{\mathbf{k}}^{\mathrm{eq}}$ is the equilibrium distribution function and we have chosen to work in units where the equilibrium temperature is $T_{\mathrm{eq}} = 1$. 
The evolution of the distribution function is then determined by the set of moments $c_n(\tau)$. Due to the conservation laws, $c_0 = 1$ and $c_1 = 0$ for all times.
Unlike the case of a constant cross section presented in \cite{Denicol:2014tha,Bazow:2015dha}, an exact solution is not known in this case, but significant progress can be made analytically. In particular, it was shown that the moments obey an infinite set of coupled ordinary differential equations that are derived in \cite{Mullins:2022fbx}. After truncation, these ordinary differential equations can then be solved numerically to find $c_n(\tau)$ in a robust manner, and correspondingly the evolution of the distribution function, as shown in \cite{Mullins:2022fbx}. The non-equilibrium jet-quenching tensor can then be determined using Eq.~\eqref{eq:Q_final_phi4} by determining $m_\varphi^2$ using the numerical solution reconstructed from the moments.

\begin{figure*}
    \centering
    \includegraphics[width=0.8\linewidth]{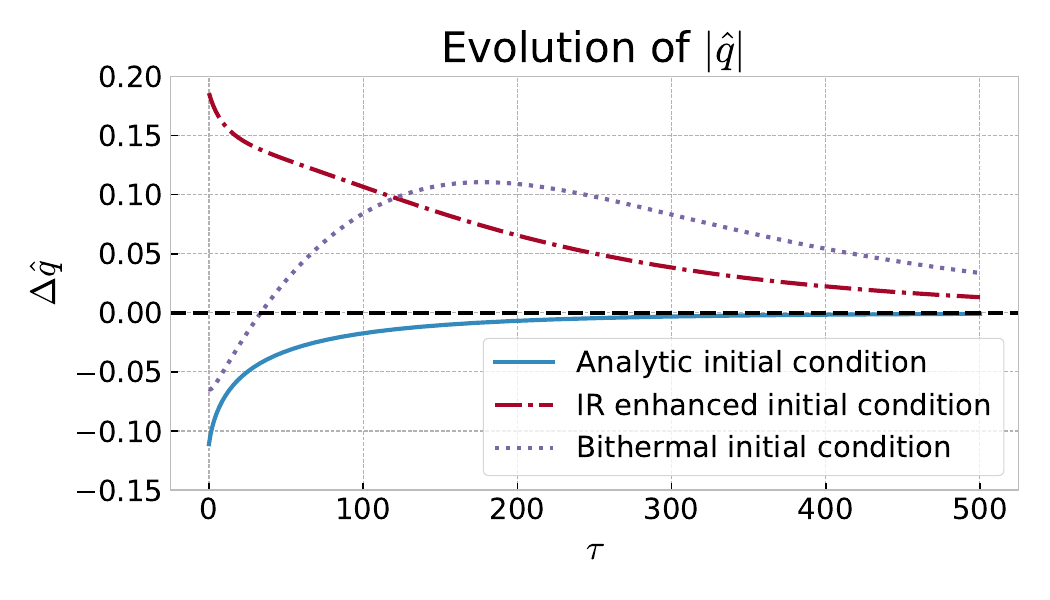}
    \caption{Variation of momentum broadening from equilibrium, as defined in Eq.~\eqref{Eq:Delta_q} for various initial conditions. The blue solid curve is the initial condition defined in Eq.~\eqref{Eq:analytic_ic}, which is under-populated. The red dotted-dashed curve is the initial condition with a Gaussian IR enhancement defined in Eq.~\eqref{Eq:Gaussian_IR_enhanced} with $A=1$ and $\sigma = 1/\sqrt{2}$. Finally, the purple dashed curve is the bithermal initial condition defined in Eq.~\eqref{Eq:bithermal_ic} with $T_1 = 2 T_{\mathrm{eq}}$, $T_2 = T_{\mathrm{eq}} / 3$, and $A = 0.35$.}
    \label{fig:q_hat_phi4}
\end{figure*} 

To compare the non-equilibrium value of $\hat{q}^{\mu\nu}$ to its equilibrium value, it is useful to isolate the magnitude of Eq.~\eqref{eq:Q_final_phi4} by defining
\begin{equation}
    |\hat{q}| = \frac{g m_{\varphi}^2}{6 E_{\mathbf{k}}} . 
\end{equation}
The only nontrivial aspect of $|\hat{q}|$ is the in-medium mass scale, determined by $m_{\varphi}^2$, which can easily be obtained in terms of the moments $c_n(\tau)$, see \cite{Mullins:2022fbx}. 
Then, the magnitude of non-equilibrium corrections can be analyzed by considering the quantity 
\begin{equation} \label{Eq:Delta_q}
    \Delta \hat{q} = \frac{|\hat{q}| - |\hat{q}|_{\mathrm{eq}}}{|\hat{q}|_{\mathrm{eq}}} , 
\end{equation}
which has a range $\Delta \hat{q} \in [-1, \infty)$. This expression is plotted as a function of $\tau = \lambda n_0 t$, with $n_0$ being the initial density, in Fig.~\ref{fig:q_hat_phi4} for various initial conditions. The first initial condition is defined through the moments as
\begin{equation} \label{Eq:analytic_ic}
    c_n(0) = \frac{1-n}{4^n} \quad \rightarrow \quad f_{\mathbf{k}}(0) = \frac{256}{243} k \, e^{-4k / 3} .
\end{equation}
In the case of a constant cross section, there exists an exact analytical solution with these initial conditions \cite{Denicol:2014tha, Bazow:2015dha}, and a numerical solution with these initial conditions was studied in \cite{Mullins:2022fbx}. We next consider a Gaussian IR bump, which is defined by 
\begin{equation} \label{Eq:Gaussian_IR_enhanced}
    f_{\mathbf{k}}(0) = f_{\mathbf{k}}^{\mathrm{eq}} \left( 1 + A e^{-k^2 / 2 \sigma^2} \right) . 
\end{equation}
This initial condition reduces to the equilibrium value rapidly over momentum scales larger than $\sigma$, but it is over-occupied in the IR by an amount characterized by $A$. This provides a simple qualitative model for the IR over-occupied distribution expected at early times in heavy-ion collisions \cite{Schlichting:2019abc,Berges:2020fwq,Lappi:2006fp,Gelis:2006dv,Keegan:2016cpi}.
The final initial condition we consider is a bithermal distribution defined by 
\begin{equation} \label{Eq:bithermal_ic}
    f_{\mathbf{k}}(0) = A e^{-k / T_1} + (1-A) e^{-k / T_2} . 
\end{equation}
This sort of distribution is used in plasma physics \cite{d_r_nicholson_introduction_1983}, and astrophysical modeling of black hole accretion disks \cite{Shapiro1976ApJ}. For the present work, it is interesting because it allows the mass scale to be close to the equilibrium value, even when the system is far from equilibrium. 

The two initial conditions presented above provide simple examples of an under-populated and over-populated distribution function, respectively. This can be seen in Fig.~\ref{fig:q_hat_phi4}, which shows the evolution of $\Delta \hat{q}$ over time. In the analytic initial condition of Eq.~\eqref{Eq:analytic_ic} the jet quenching parameter approaches equilibrium from below, while in the Gaussian enhanced initial condition of Eq.~\eqref{Eq:Gaussian_IR_enhanced} it approaches equilibrium from above. This makes it clear that, depending on the specific structure of the out-of-equilibrium distribution, the medium can, in some cases, lead to less momentum broadening than the equilibrium case, highlighting that jet quenching can be suppressed depending on the far-from-equilibrium scenarios being considered. We note that both distributions relax toward the same equilibrium Boltzmann limit at late times. Therefore, differences in $\delta \hat{q}$ arise only from the initial non-equilibrium initial condition. 

The bithermal initial condition of Eq.~\eqref{Eq:bithermal_ic} has the most striking features of the initial conditions. The evolution of this distribution shows non-monotonic behavior, starting with a reduction of the jet quenching parameter but transitioning to an increase at later times that slowly decreases as the system equilibrates. This nicely shows that the evolution of the jet quenching parameter can have nontrivial structure depending strongly on the initial condition in out of equilibrium media.

\section{Summary and Discussion}
\label{Sec:Conclusions}

In this work, we proposed a covariant generalization of the jet transport coefficient $\hat q$ by promoting it from a single scalar to a Lorentz tensor, $\hat q^{\mu\nu}$. The need for this tensorial generalization is that momentum broadening is a diffusion process in momentum space, and in a relativistic, flowing, or out-of-equilibrium medium there is no reason for that diffusion to be fully captured by one single quantity. The tensor $\hat q^{\mu\nu}$ organizes the same underlying microscopic physics in a frame-covariant way, making explicit which components are redundant in equilibrium and which become genuinely independent once isotropy and time-independence are lost. In particular, the tensor formulation gives direct access not only to transverse momentum broadening, but also to energy broadening and to energy--momentum correlations encoded in the off-diagonal components---information that is invisible in the standard scalar characterization.

We developed this framework within a leading-order elastic (Boltzmann/Fokker--Planck) description and showed that $\hat q^{\mu\nu}$ naturally admits a diffusion interpretation: for a narrow, high-energy jet it governs the growth of the jet momentum covariance per unit path length. This perspective immediately implies diffusion-type constraints: because $\hat q^{\mu\nu}$ is constructed as the second moment of a positive scattering kernel, it must be symmetric and positive semidefinite, and in the local rest frame the spatial matrix $\hat q^{ij}$ must have nonnegative eigenvalues. These constraints provide a clean diagnostic for when a Markovian kinetic description of broadening is self-consistent, and they suggest a systematic way to test or constrain phenomenological models that attempt to parametrize jet broadening beyond equilibrium.

To illustrate the formalism in a controlled setting, we computed $\hat q^{\mu\nu}$ explicitly in tree-level massless $\lambda\varphi^4$ theory for isotropic but time-dependent out-of-equilibrium states. We performed the calculation both with Bose--Einstein statistics and in the classical (Boltzmann) limit. In the regime where the jet momentum is large compared to the characteristic medium momentum scale, we find that quantum statistical contributions are subleading and the resulting $\hat q^{\mu\nu}$ is well approximated by its classical counterpart. This motivates using the classical description as a reliable and analytically tractable approximation for studying how non-equilibrium evolution feeds into broadening in this toy model. In particular, we find that the leading out-of-equilibrium correction enters through the in-medium mass scale $m_\varphi^2$, which acts as a screening-like medium scale in the present context. Solving the classical Boltzmann equation within a method-of-moments framework, we determined the time dependence of $\hat q^{\mu\nu}(t)$ during relaxation and showed that out-of-equilibrium corrections can either enhance or reduce the broadening relative to equilibrium, depending on the initial distribution function.

There are several natural directions to pursue. First, while we constructed $\hat q^{\mu\nu}$ in kinetic theory, it would be valuable to formulate the same covariant diffusion tensor starting from other standard definitions of $\hat q$, such as field-theoretic correlator \cite{Majumder:2012sh,Casalderrey-Solana:2007ahi,Sadofyev:2021ohn} and Wilson-line formulations \cite{Liu:2006ug,DEramo:2010wup,Panero:2013pla}, in order to connect directly to nonperturbative approaches and to clarify scheme and operator definitions in the covariant setting. Second, the most phenomenologically relevant extension is to compute $\hat q^{\mu\nu}$ in QCD effective kinetic theory \cite{Caron-Huot:2008zna,Ghiglieri:2015ala,Boguslavski:2023alu,Boguslavski:2023waw} and quantify its behavior in far-from-equilibrium plasmas with strong flow and anisotropy. Finally, since heavy-ion phenomenology often models the medium with relativistic hydrodynamics after hydrodynamization, it will be interesting to incorporate $\hat q^{\mu\nu}$ into jet quenching simulations on top of hydrodynamic backgrounds, where spacetime-dependent flow and dissipative stresses provide the additional tensors that can feed into the irreducible decomposition of $\hat q^{\mu\nu}$. In this context, we hope that the covariant organization provided by $\hat q^{\mu\nu}$ can serve as a bridge between first-principles microscopic descriptions and more realistic dynamical modeling of jet--medium interactions out of equilibrium.

\section{Acknowledgments}
We would like to thank Gabriel Rocha and Caio Brito for interesting discussions on these calculations. I.D. and J.N. were
partly supported by the US-DOE Nuclear Science Grant No. DE-SC0023861 and by
the National Science Foundation (NSF) within the framework of the MUSES collaboration, under grant number
OAC-2103680.
N.M. was partially
supported by U.S. Department of Energy, Office of Nuclear Physics, Contract DE-FG02-03ER41260.
Any opinions, findings, and conclusions
or recommendations expressed in this material are those
of the author(s) and do not necessarily reflect the views of
the National Science Foundation or U.S. Department of Energy.
\bibliographystyle{apsrev4-2}
\bibliography{scalar_qhat_v3}
\end{document}